\documentclass[11pt,english]{article}
\usepackage[round]{natbib}

\usepackage{pdflscape}
\usepackage{color}
\usepackage{latexsym,amsmath,amssymb,graphicx}
\usepackage{amsthm}
\usepackage{makeidx}
\usepackage{enumerate}
\usepackage{dsfont}

\usepackage[round]{natbib}
\usepackage{amsfonts}
\usepackage{verbatim}
\usepackage{rotating}

\usepackage{hyperref}

\newtheorem{lemma}{Lemma}[section]

\begin{document}

\title{Optimal dose calibration in radiotherapy}

\author{Jes\'us L\'opez-Fidalgo$^{1}$ and Mariano Amo-Salas$^2$}
\date{}
\maketitle
\begin{center}
{\scriptsize $^1$University of Navarra, ICS, Unit of Data Science\\ $^2$Department of Mathematics, University of Castilla-La Mancha}
\end{center}


\begin{abstract}
In this paper, the tools provided by the theory of Optimal Experimental Design are applied to a nonlinear calibration model. This is motivated by the need of estimating radiation doses using radiochromic films for radiotherapy purposes. The calibration model is in this case nonlinear and the explanatory variable cannot be worked out explicitly from the model. In this case an experimental design has to be found on the dependent variable. For that, the inverse function theorem will be used to obtain an information matrix to be optimized. Optimal designs on the response variable are computed from two different perspectives, first for fitting the model and estimating each of the parameters and then for predicting the proper dose to be applied to the patient. While the first is a common point of view in a general context of the Optimal Experimental Design, the latter is actually the main objective of the calibration problem for the practitioners and algorithms for computing these optimal designs are also provided.

The optimal designs obtained have just three different points in their support, but practitioners usually demand for more support points. Thus, a methodology for computing space-filling designs is also provided when the support points are forced to follow some mathematical rule, such as arithmetic or geometric sequences. Cross efficiencies of all these designs are computed in order to show their ability for different goals. 

\end{abstract}

\section{Introduction}

Calibration models are used in many scientific and industrial fields. They have been studied widely, e.g. by \cite{Osb91}. It means a different perspective from a standard experimental regression model. In particular, the calibration process is made in two steps. First, for known values of the explanatory variable, the response is measured and the parameters of the model are fitted. Then, on a second stage, in order to calibrate a particular value of the explanatory variable, the response is computed using the inverse function of the model and after that the right value of the explanatory variable to be used is predicted. Thus, while for a standard regression model, generally the main concern is the estimation of the parameters of the model or the prediction of the response at some values of the explanatory variable, for a calibration model the main concern is the prediction of the explanatory variable in order to get a desired target for the response.

Optimal designs for calibration models have been rarely considered in the literature. \cite{Kit92}, provided a procedure in a simple case when the explicit expression of the inverse model can be obtained and the outline becomes traditional.  \cite{Fra04} computed optimal designs for inverse prediction in calibration models and presented two criteria, G$_I-$optimality and V$_I-$optimality. The aims of these criteria are the same than G- and V-optimality, but for the case of inverse prediction. \cite{Biederman11} considered a similar problem for indirect observations through a second variable and shows how a uniform design performs quite well for this purpose. Finally, \cite{Amo16} presented a previous work in this field, which is detailed and extended in this paper from the perspective of calibration. Thus, the nominal values considered by \cite{Amo16} allow obtaining a closed--form expression for the inverted model. In this work, this constraint is removed and a more general framework is presented. Then, the aim of this paper is the study of nonlinear models where the explanatory variable is expressed as a function of the dependent variable and this function has not a closed--form for its inverse. This study is presented from two perspectives, firstly it is focused on the estimation of the parameters of the model and then on the final goal is prediction of the explanatory variable. 

One of the fields where the calibration has an important role is dosimetry.  The use of digital radiographs has been a turning point in dosimetry. In particular, radiochromic films are very popular nowadays because of their near tissue equivalence, weak energy dependence and high spatial resolution. In this area, calibration is frequently used to determine the right dose in radiotherapy for a specific patient. A film is first irradiated at known doses for building a calibration table, which will be used to fit a parametric model, where now the dose plays the role of the dependent variable.

\cite{Ram13} used the procedure described ahead. The radiochromic films were scanned twice. The first scanning was made when a pack of films arrived and the second 24 hours after being irradiated. With the two recorded images the optical density, $netOD$, is calculated as the base 10 logarithm of the ratio between the means of the pixel values before ($PV_{0}$) and after ($PV$) the irradiation. They used patterns formed by 12 squares of $4 \times 4\, cm^2$ irradiated at different doses. This size is assumed enough to ensure the lateral electronic equilibrium for the beam under consideration. A resolution of 72 pixels per inch without color correction and with 48-bit pixel depth was used for the measurements. The pixel values were read at the center of every square. Then the mean and the standard error were calculated for each square.

To adjust the results to the calibration table the following model has been frequently used:
\begin{eqnarray*}
netOD=\eta(Dose,\theta)+ \varepsilon,
\end{eqnarray*}
where the error $\varepsilon$ will be assumed normally distributed with mean zero and constant variance, $\sigma^2$. The expression of the function $\eta(Dose,\theta)$ is unknown but the mathematical formula of the inverse is a known function,
\begin{eqnarray}
\eta^{-1}(Dose, \theta)=\mu(netOD,\theta) = \alpha \ netOD + \beta \ netOD ^{ \gamma }, \; Dose\in \mathcal{X}_{Dose}=[0,B], \label{e.radmodel}
\end{eqnarray}
where $\theta=(\alpha,\beta,\gamma)^T$ are unknown parameters to be estimated using the Least Squares procedure (LSE), actually the Maximum Likelihood Estimates (MLE) in the case of normality.

In this paper a case study from \cite{Rein12} will be used to illustrate the procedure. They considered the case of dose verification in highly conformal radiation therapy taking advantage of the high spatial resolution offered by radiochromic films such as Gafchromic EBT, EBT2 or the new generation of these films EBT3. \cite{mH19} computed optimal designs for calibration dosimetry models in a different way.

This work is structured as follows, in Section \ref{s.oed} the inverse function theorem is used to obtain the expression of the Fisher Information Matrix (FIM), which is needed for computing optimal designs. This work considers both, estimating the parameters of the model (Section \ref{s.est}) and calibrating the explanatory variable of the radiation dose (Section \ref{sec:cal}) and for both situations optimal designs are computed using a study case from \cite{Rein12}. Moreover, optimal space--filling designs are computed in both cases for reaching the requirements of the common practice.

\section{Optimal experimental design for calibration}\label{s.oed}
Let a general model be

\begin{equation}
y=\eta(x,\theta)+\varepsilon, \quad \varepsilon\sim N(0,\sigma), \label{e.model}
\end{equation}
where $y$ is the dependent variable, $x$ is the explanatory variable, $\theta$ is the vector of parameters of the model and $\eta(x,\theta)$ is an unknown function with $\mu(y,\theta)=\eta^{-1}(x,\theta)$ explicitly known. The challenge here is to find optimal designs for the explanatory variable when the expression of the function $\eta(x,\theta)$ is unknown and possibly nonlinear in the parameters.

An {\it exact experimental design of size $n$} consists of a collection of points $x_i, \, i=1, ..., n$, in a given compact {\it design space}, $\mathcal{X}$. Some of these points may be repeated and a probability measure can be defined assigning to each different point the proportion of times it appears in the design. This leads to the idea of extending the definition of experimental design to any probability measure ({\it approximate design}). From the optimal experimental design point of view we can restrict the search to discrete designs of the type
$$ \xi = \bigg \{ \begin{array}{cccc}  x_{1} & x_{2} & ... & x_{k}\\ p_1 & p_2 & ... & p_k \end{array} \bigg \},$$
where $x_i, \ i=1, ..., k$ are the support points and $\xi(x_i)=p_i$ is the proportion of experiments to be made at point $x_i$. Thus, $p_i\geq 0 $ and $\sum_{i=1}^{k} p_i=1$.

For the exponential family of distributions the FIM of a design $\xi$ is given by
\begin{eqnarray}
M(\xi,\theta)=\sum_{x\in\mathcal{X}}I(x,\theta)\xi(x), \label{e.fim}
\end{eqnarray}
where $I(x,\theta)=\frac{\partial\eta(x,\theta)}{\partial\theta} \frac{\partial\eta(x,\theta)}{\partial\theta^T}$ is the FIM at a particular point $x$. It is evaluated at some nominal value of $\theta$. This is actually the FIM of a linear model with regressors $\frac{\partial\eta(x,\theta)}{\partial\theta}$. The nominal value usually represents the best guess for the parameters vector $\theta$ at the beginning of the experiment.

It can be proved that the inverse of this matrix is asymptotically proportional to the covariance matrix of the parameter estimators. An optimality design criterion, $\Phi[M(\xi,\theta)]$, aims to minimize the covariance matrix in some sense and therefore the inverse of the information matrix. For simplicity $\Phi(\xi)$ will be used instead of $\Phi[M(\xi,\theta)]$. In this paper two popular criteria, D-- and c-optimality, as well as two calibration-oriented criteria, G$_I-$ and V$_I-$optimality, will be considered.

The D-optimality criterion minimizes the volume of the confidence ellipsoid of the parameters and it is given by $\Phi_{D}(\xi)=\det M^{-1/m}(\xi,\theta)$, where $m$ is the number of parameters in the model. The c-optimality criterion is used to estimate a linear combination of the parameters, say $c^{T}\theta$, and it is defined by $\Phi_{c}(\xi)=c^{T}M^{-}(\xi,\theta)c$, where the superscript $^{-}$ stands for the generalized inverse class of the matrix. Although the generalized inverse is unique only for nonsingular matrices the value of $c^{T}M(\xi,\theta)^{-}c$ is invariant for any member of the generalized inverse class if and only if $c^{T}\theta$ is estimable with the design $\xi$.

\cite{Fra04} computed optimal designs for inverse prediction in calibration models and presented two criteria, G$_I-$ and V$_I-$optimality. The aims of these criteria are the same that $G-$ and $V-$optimality, that is minimizing the maximum and the average prediction variance respectively, but when the interest is in inverse prediction. These criteria will be detailed in Section \ref{sec:cal} as well as the algorithms for computing the optimal designs. These criterion functions are convex and non-increasing. A design that minimizes one of these functions, say $\Phi$, over all the designs defined on $\mathcal{X}$ is called a $\Phi-$optimal design, or more specifically, a D-, c-, G$_I-$ or V$_I-$optimal design.

The goodness of a design, $\xi$, is measured by its efficiency, defined by
$$\mbox{eff}_{\Phi}(\xi)=\frac{\Phi(\xi^{*})}{\Phi(\xi)},$$
where $\xi^{*}$ is the $\Phi-$optimal design. 

This efficiency can be multiplied by 100 and be reported in percentage. If the function has a homogeneity property there is a practical statistical interpretation. Thus, if the efficiency of a design is $50\%$ this means that the design needs to double the total number of observations to perform as well as the optimal design.

In order to check the optimality of a design the General Equivalence Theorem (GET) can be used (\citealp{Kie60, Whi73}) for a more general version. This theorem is valid for approximate designs and convex criteria. It is quite useful also for building efficient algorithms for computing optimal designs.  Let $\psi(x,\xi)$ be the Frechet directional derivative in the direction of a one-point design at $x$,
\begin{eqnarray*}
\psi(x,\xi)=\lim_{\varepsilon \rightarrow 0^+}\frac{\Phi ((1-\varepsilon)M(\xi,\theta)+\varepsilon I(x,\theta))-\Phi (M(\xi,\theta))}{\varepsilon}.
\end{eqnarray*}

 This function is frequently called the sensitivity function.
The GET states that under some conditions of the criterion function, $\psi(x,\xi)$ achieves its minimum value, zero, at the support points of the optimal design.

This theorem provides also a bound for the $\Phi$--efficiency of a design, $\xi$,
$$ \mbox{eff}_{\Phi}(\xi)\geq 1 + \frac{\min_{x} \psi(x,\xi)}{\Phi(\xi)}.$$
For D-optimality $\psi(x,\xi^*)=m-\frac{\partial\eta(x,\theta)}{\partial\theta^T} M^{-1}(\xi,\theta)\frac{\partial\eta(x,\theta)}{\partial\theta}$.

For c-optimality the Elfving's graphic method (\citealp{Elf52}) can be used to construct the optimal design and this will not be needed. The G$_I-$optimal criterion is not differentiable and for V$_I$-optimality the sensitivity function will be given in Section \ref{sec:cal}.

More details on the theory of optimal experimental designs may be found, e.g., by \cite{Paz86,Fed97,Atk07}.

\subsection{Inverse function theorem for computing the FIM}\label{sec.inv}

The experiments are designed for the explanatory variable, $x$, which is assumed under the control of the experimenter. However, in this work it is considered that  $\eta(x,\theta)$ is unknown and  invertible within the design space. Nevertheless, the expression of the inverse with respect to $x$, $\mu(y,\theta)=\eta^{-1}(x,\theta)$, is known. Therefore the FIM, which is given by (\ref{e.fim}), is defined in terms of $y$ instead of $x$. In particular, for a specific point the FIM is

$$I(x,\theta)=\frac{\partial\eta(x,\theta)}{\partial\theta} \frac{\partial\eta(x,\theta)}{\partial\theta^T}.$$

We can calculate the FIM in terms of the response variable $y$ through the inverse function theorem and the chain rule for differentiating composed functions. In particular, differentiating the equation
$$x=\mu(y,\theta)=\mu(\eta(x,\theta),\theta),$$
we obtain
$$ 0=\left( \frac{\partial\mu (y,\theta)}{\partial y} \right)_{y=\eta(x,\theta)} \frac{\partial\eta(x,\theta)}{\partial\theta} + \left(\frac{\partial\mu (y,\theta)}{\partial\theta}\right)_{y=\eta(x,\theta)}.$$
Then
\begin{eqnarray}
 \frac{\partial\eta(x,\theta)}{\partial\theta} = -\left(\frac{\partial\mu(y,\theta)}{\partial y} \right)_{y=\eta(x,\theta)}^{-1} \left(\frac{\partial\mu (y,\theta)}{\partial\theta}\right)_{y=\eta(x,\theta)}. \label{e.transfim}
 \end{eqnarray}
For simplicity of notation the last expression will be called $f(x)$.

This result allows the computation of the FIM and therefore optimal designs on $x$ may be obtained. This is the same model to be used for designing variable $y$ in the inverse model being heteroscedastic instead of homoscedastic, with  variance
\begin{eqnarray*}
\left(\frac{\partial\mu(y,\theta)}{\partial y} \right)^2.
\end{eqnarray*}

This is meaningful since if the response is considered as the mean model plus some error with constant variance, then the mean model for the explanatory variable could be approximated by the inverse of the original mean plus a different error, now with a non--constant variance coming from the transformation of the model.

\section{Designs for best fitting the model}\label{s.est}

The model proposed by \cite{Rein12} is being considered for the case study. In this model, the function $\eta(Dose,\theta)$ is unknown but its inverse is known and defined by Equation (\ref{e.radmodel}). Using Equation (\ref{e.transfim}), we obtain

\[f(netOD)=\frac{\partial\eta(Dose,\theta)}{\partial\theta}=\frac{-1}{\alpha_{0}+\beta_{0}\gamma_{0}\eta(Dose,\theta)^{\gamma_{0}-1}} \left( \begin{array}{ccc}
\eta(Dose,\theta)  \\
\eta(Dose,\theta)^{\gamma_{0}}\\
\beta_{0} \eta(Dose,\theta)^{\gamma_{0}}\log(\eta(Dose,\theta))\end{array} \right),\]
where $\alpha_{0}, \beta_{0}, \gamma_{0}$ are some nominal values assumed for the parameters to compute the optimal design. The function $f(netOD)$ is considered here as a function of $netOD$ since mathematical expression of $\eta(Dose,\theta)$ is not known and it will be considered as the inverse of $\mu(netOD,\theta)$. 
Thus, the FIM for a design  $\xi$ is

$$ M(\xi; \theta_{0})  = \sum_{i}\xi(Dose_i)I(Dose_i, \theta_{0}),$$
where $\theta_{0}^T=(\alpha_{0}, \beta_{0}, \gamma_{0})$, $netOD$ has been replaced by $\eta(Dose,\theta)$ and

\begin{equation}
I(Dose, \theta_{0})=\frac{1}{(\alpha_{0}+\beta_{0}\gamma_{0}\eta^{\gamma_{0}-1})^2}
\left( \begin{array}{ccc}
\eta &  \eta^{\gamma_{0}+1} & \beta_{0} \eta^{\gamma_{0}+1}\log(\eta) \\
\eta^{\gamma_{0}+1} & \eta^{2 \gamma_{0}}& \beta_{0} \eta^{2 \gamma_{0}}\log(\eta)\\
\beta_{0} \eta^{\gamma_{0}+1}\log(\eta) & \beta_{0} \eta^{2 \gamma_{0}}\log(\eta)& \beta_{0}^2 \eta^{2\gamma_{0}}\log^2(\eta)\end{array} \right),\label{e.fimrad}
\end{equation}
where $\eta$ denotes $\eta(Dose,\theta)$ for simplicity of notation.

Although the optimal design is computed on $Dose$ the function of the original model $\eta(Dose,\theta)$ is unknown, then this function is replaced by $netOD$ in Equation (\ref{e.fimrad}). Using the results of \cite{Rein12}, the estimated parameters for radiochromic new generation films  EBT3 (in particular the F06110902 film lot and radiation type Proton, Table I) are being considered as nominal values: $ \alpha_{0} = 8.32 $, $ \beta_{0} =49.91 $ and $ \gamma_ {0} = 2.6$. Just for some values of $\gamma_ {0}$ the inverse function can be computed analytically, otherwise it has to be computed numerically when needed, what makes the problem much more demanding from a computational point of view. The design space on the $netOD$, $ \mathcal{X}_{netOD} = [0,b]=[0,0.45]$, corresponds to the design space for the dose, $ \mathcal{X}_{Dose} =[0,B]= [0,10.00]$.

\subsection{D-optimal designs}

\begin{figure}[h]
\begin{center}
\includegraphics[scale=0.26]{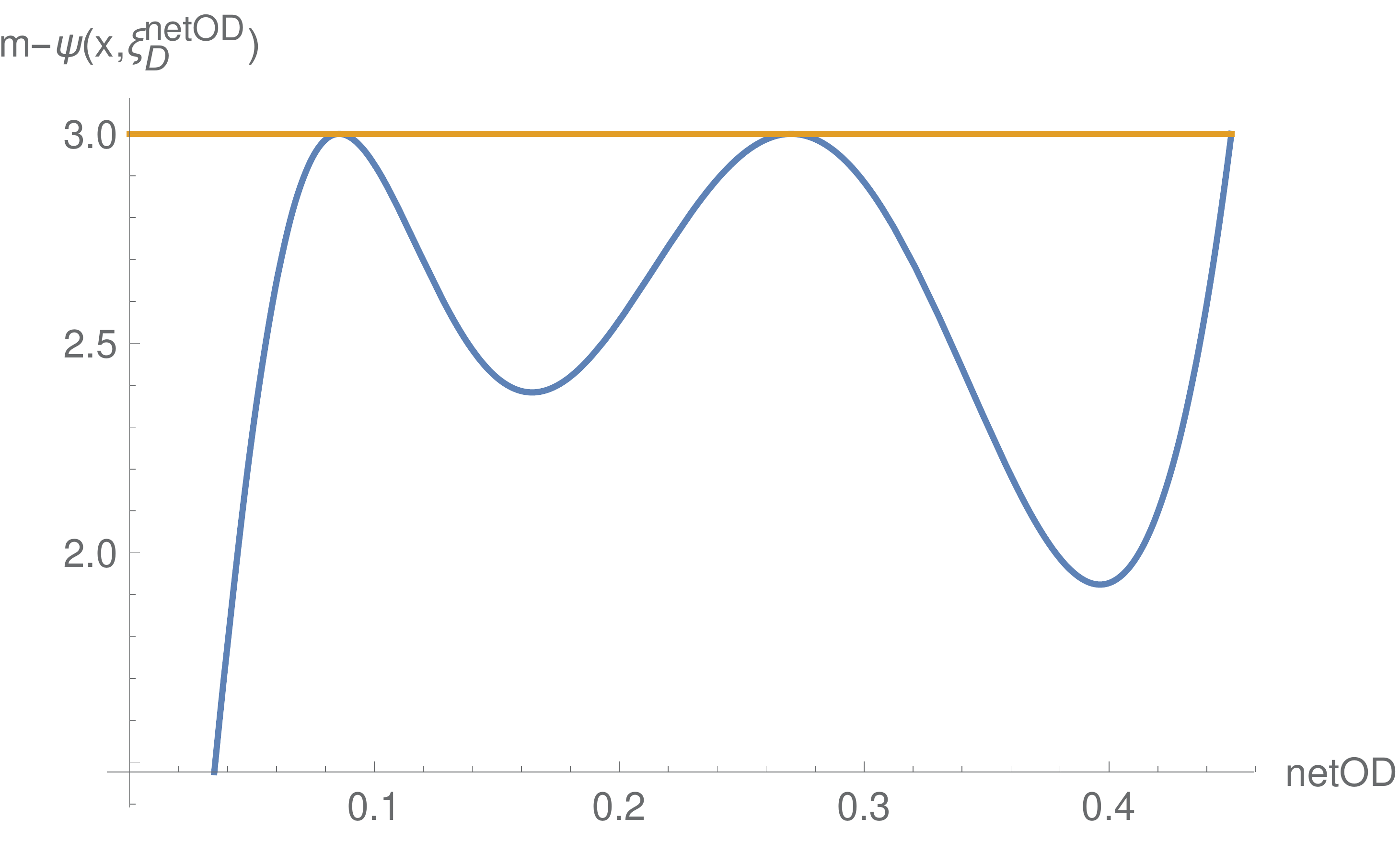}\includegraphics[scale=0.26]{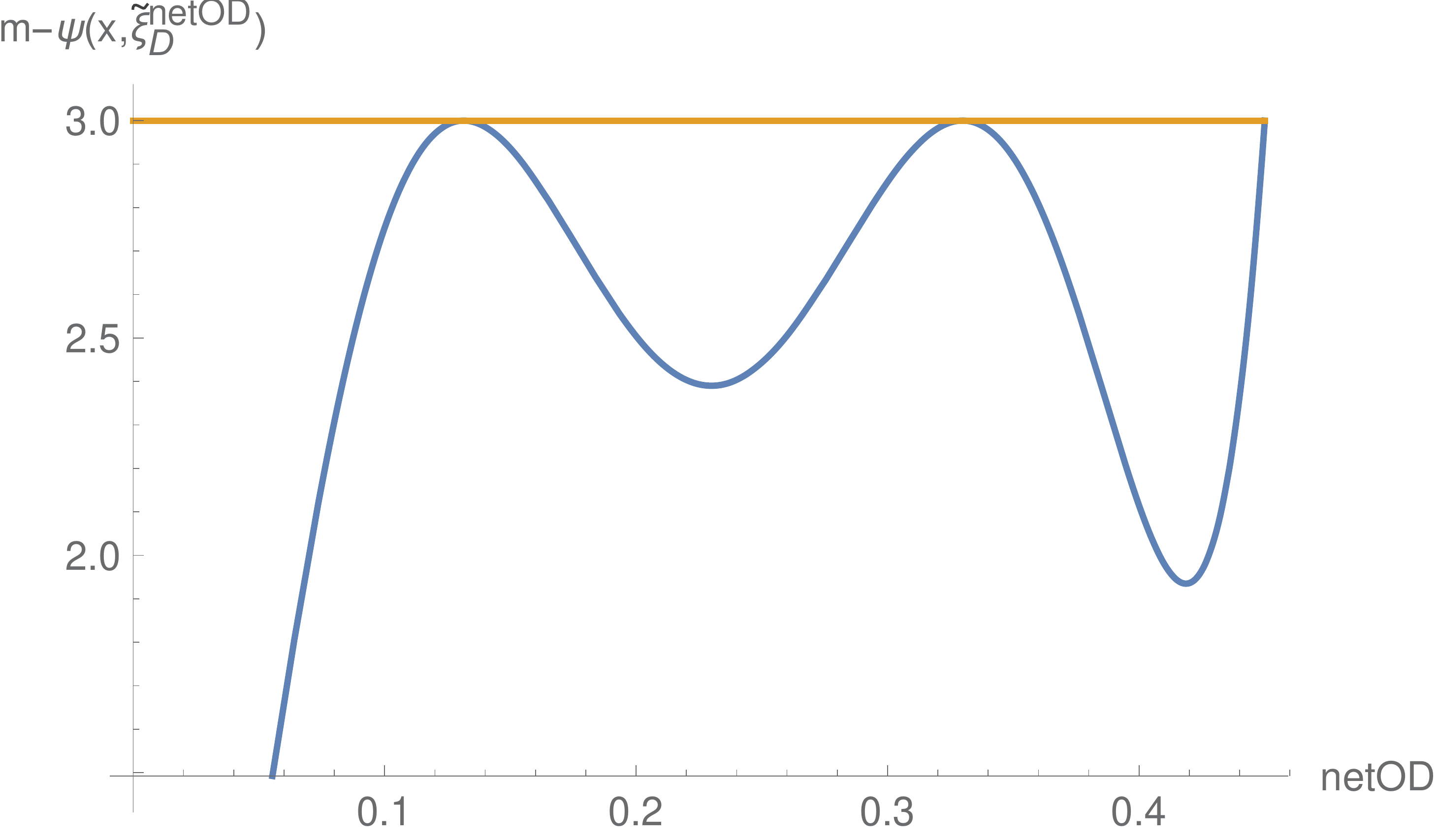}
\end{center}
\caption{Sensitivity function for $\xi_{D}^{netOD}$ (left) and $\tilde{\xi}_{D}^{netOD}$ (right) designs.}\label{f.sens1}
 \end{figure}

In order to compute the $D-$optimal design and due to the number of parameters, a three--point design with equal weights in the support points will be assumed, say $Dose_1$, $Dose_2$ and  $Dose_3$ with weights 1/3.
The D-optimal is computed on the variable $netOD$ using matrix (\ref{e.fimrad}) and then transformed into a design on the variable $Dose$. The determinant of the information matrix for a general three--point design supported on $netOD_1$, $netOD_2$ and $netOD_3$ with weights 1/3 at each point is computed and maximized in the interval $ \mathcal{X}_{netOD} = [0,b]=[0,0.45].$ The obtained design, $\xi_{D}^{netOD}$, shown in Table \ref{t.new}, is actually D-optimal. Figure \ref{f.sens1} (left) shows

$$m-\psi(x,\xi_{D}^{netOD})=\frac{\partial\eta(Dose,\theta)}{\partial\theta^T} M^{-1}(\xi_{D}^{netOD},\theta)\frac{\partial\eta(Dose,\theta)}{\partial\theta}$$ 
for the design obtained using the transformation of the theorem of the inverse function. It is lower than the number of parameters, $m=3$, and therefore the sensitivity function, $\psi(x,\xi_{D}^{netOD})\geq 0$. The equivalence theorem states this design is actually D-optimal.

Transforming the three points through the function (\ref{e.radmodel}), with the previous nominal values of the parameters,
$$  Dose = 8.32  \times netOD + 49.91  \times netOD ^{ 2.6 } ,$$
the D-optimal design on variable $Dose$, $\xi_{D}^{Dose}$, is shown in Table \ref{t.new}.

Now a design for $netOD$ is computed in the usual way for the function $\mu(netOD,\theta)$ in order to compare it with the previous one and check the loss of efficiency. This is the wrong way of computing it since the roles of the independent and the dependent variables are actually exchanged. In particular the variance term is considered constant in this way missing the heteroscedasticity induced by inverting the model. Then, $netOD$ is considered as the explanatory variable and after computing the D-optimal design for $netOD$, we obtain the transformed design for $Dose$. That is, we are assuming $\mu(netOD,\theta)$ as the function of the original model. Using the previous nominal values, the D-optimal design, say $\tilde{\xi}_{D}^{netOD}$ is given in Table \ref{t.new}.

Figure \ref{f.sens1} (right) shows that 

$$m-\psi(x,\tilde{\xi}_{D}^{netOD})=\frac{\partial\mu(netOD,\theta)}{\partial\theta^T} M^{-1}(\tilde{\xi}_{D}^{netOD},\theta)\frac{\partial\mu(netOD,\theta)}{\partial\theta}$$ 
is lower than the number of parameters, $m=3$, and therefore the sensitivity function is greater than or equal to zero. The equivalence theorem states this design is actually D-optimal. At this point a design for the response, $Dose$, can be obtained by transforming again the design points using the equation from the model $\mu(netOD,\theta)$ (Table \ref{t.new}).

Apparently this design is quite different from the correct one, e.g. the first and the second points are sensitively larger than the originals. But the efficiency provides better information to compare this design with respect to the right one, $\xi_{D}^{Dose}$,

$$  \mbox{eff}_{D}(\tilde{\xi}_{D}^{Dose})=\left(\frac{ \Phi_D(\xi_{D}^{Dose})}{ \Phi_D(\tilde{\xi}_{D}^{Dose})} \right)^{\frac{1}{3}}=0.868,$$
which means, in this particular case, a moderate loss of efficiency of a little less than 15\%. This suggests that it is important the use of the right expression of $f(netOD)$ for computing the optimal design. In the following sections the designs obtained will be computed directly for $netOD$ using the transformed expression of $f(netOD)$ and then transforming it back to designs on $Dose$.

\subsection{$c-$optimal designs}

Frequently the interest is not in estimating all of the parameters of the model, but some linear combination. A particular case is when there is special interest in estimating just one particular parameter. For example, in the case considered here, there is special interest in $\gamma$. As mentioned above the Elfving's method is a graph procedure for calculating c-optimal designs. Although the method can be applied to any number of parameters it is not easily visualized for more than two parameters. \cite{Lop04} proposed a computational procedure for finding c-optimal designs using Elfving's method for more than two dimensions. \cite{ha08b, Ba11} have also developed these idea.

\begin{center}
\begin{figure}[h]
\begin{center}
\includegraphics[scale=0.3]{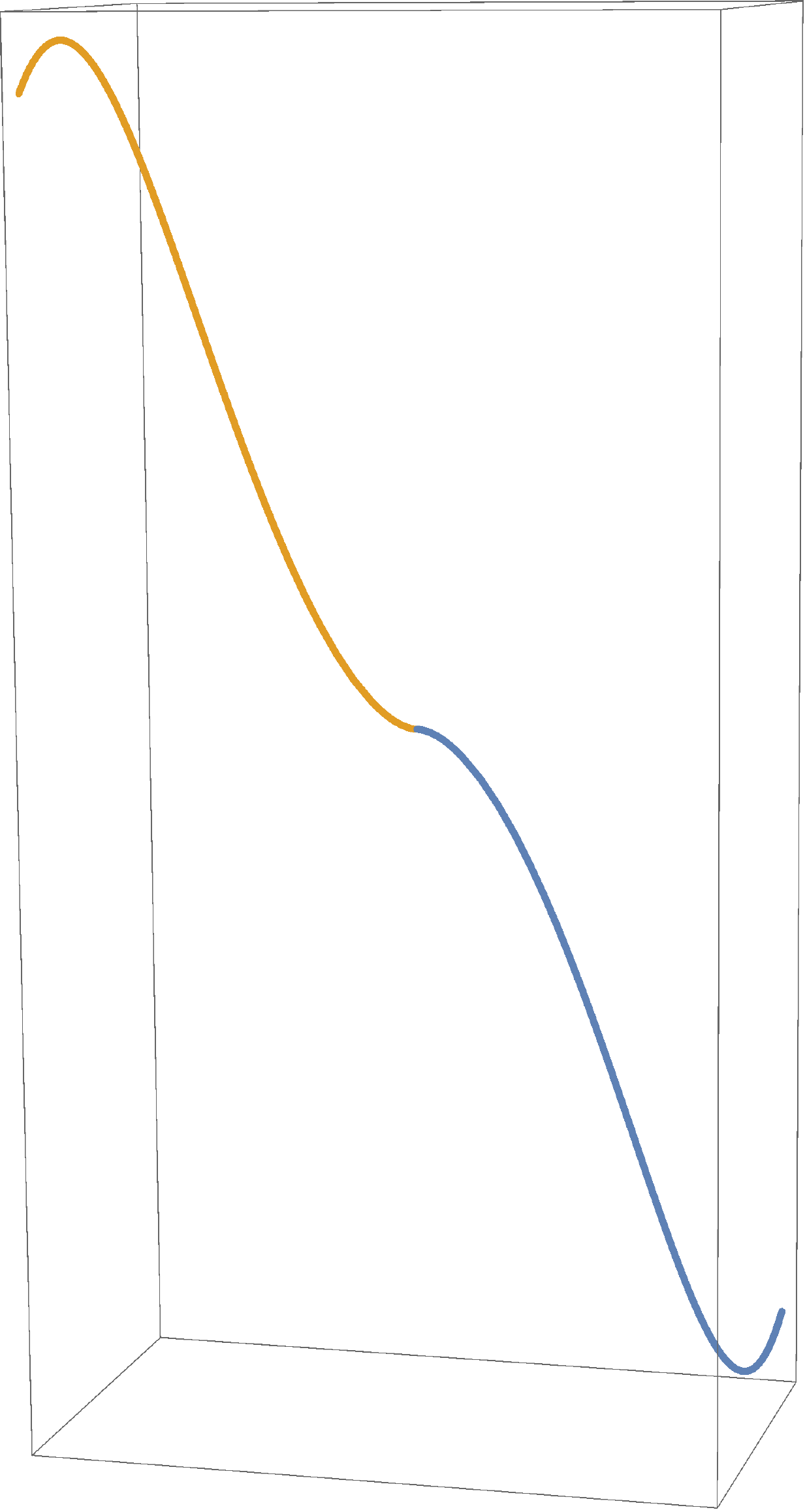}
\caption{Curves $f(\mathcal{X}_{netOD}) \cup - f(\mathcal{X}_{netOD})$ for the Elfving's locus, properly scaled to detect the shape details}
 \label{f.elf}
\end{center}
 \end{figure}
\end{center}

In the example considered in this paper there are three parameters and the Elfving locus, convex hull of $f(\mathcal{X}_{netOD}) \cup - f(\mathcal{X}_{netOD})$, where $f(x)$ comes from Equation (\ref{e.transfim}), is hard to be visualized (Figure \ref{f.elf}). Computing the intersection point of the boundary with the straight line defined by vector $c$ for the case of two parameters is rather simple, but for more parameters, even just three, it is not affordable or too difficult. Thus, the procedure detailed by \cite{Lop04} is being applied here. The idea is simple although the formalization is a bit tedious. Once one has the Elfving locus all it is needed is to find the intersection of the line defined by vector $c$ (assuming the objective is to estimate $c^T \theta$) with the boundary of this set. The two possible points are just symmetric and produce the same design. This point is a convex combination of at most $m$ points of the set $f(\mathcal{X}_{netOD}) \cup - f(\mathcal{X}_{netOD})$. Those points will be the support points of the c-optimal design and the coefficients of the convex combination will be the weights of the design.

The procedure takes into account that any of the non--null components of a generic point satisfying the conditions of the Elfving locus  can be considered as objective function. The only exceptions are the null components of vector $c$. This point has to be a convex combination of no more than $m$ support points of the set $f(\mathcal{X}_{netOD}) \cup - f(\mathcal{X}_{netOD})$. Thus, a generic point of this type depends on $m$ different points and $m-1$ different coefficients. Now they must satisfy that they are in the straight line defined by $c$, so the point must be $\rho c$, for some scalar $\rho$. This gives $m$ linear (in the coefficients) equations  with the extra $\rho$. Solving the linear system on the coefficients they will disappear. Thus, the objective function is any of the components of the point such that $c_i\neq 0$, which depends just on the $m$ points of the design. This is now a standard optimization problem with a number of algorithms and software available for computing the optimum.

In this section c-optimal designs will be computed for estimating each of the parameters in the example considered. The procedure is being explained in more detail for the computation of the c-optimal design for $\gamma$, i.e., corresponding to the vector $c^T=(0,0,1)$.

The c-optimal design will be of the form:
\[ \xi=\left\{ \begin{array}{ccc}
t & s & u  \\
1-\lambda-\delta & \lambda & \delta\end{array} \right\},\]
corresponding to a point on the boundary of the Elfving's locus as well as on the line defined by $c^T=(0,0,1)$. A point in the Elfving's locus has to be a convex combination of, at least, three points of $f(\mathcal{X}_{netOD}) \cup - f(\mathcal{X}_{netOD})$. Apart from symmetric situations there are two possibilities. Either the three points come from $f(\mathcal{X}_{netOD})$ (equivalently from $-f(\mathcal{X}_{netOD})$), that is,
$$(x_1,x_2, x_3)^T=(1-\lambda-\delta)f(t)+\lambda f(s)+\delta f(u)$$
or two come from $f(\mathcal{X}_{netOD})$ and one from $-f(\mathcal{X}_{netOD})$ (symmetrically two come from $-f(\mathcal{X}_{netOD})$ and one from $f(\mathcal{X}_{netOD})$), that is,
$$(x_1,x_2, x_3)^T=(1-\lambda-\delta)f(t)+\lambda f(s)-\delta f(u).$$
Figure \ref{f.elf} shows the point will be of the second kind. At the same time they must be on the line defined by $c^T=(0,0,1)$. Thus, for the second case,
$$(1-\lambda-\delta)f(t)+\lambda f(s)-\delta f(u)=\rho (0,0,1)^T.$$

The equations coming from the two first components give the values of $\lambda$ and $\delta$ as a function of the three points,
\begin{eqnarray*}
(1-\lambda-\delta)f_1(t)+\lambda f_1(s)-\delta f_1(u)=0,\\
(1-\lambda-\delta)f_2(t)+\lambda f_2(s)-\delta f_2(u)=0,
\end{eqnarray*}
that is
\begin{eqnarray}\nonumber
\lambda(t,s,u)&=&\frac{{f_1(u)} {f_2(t)}-{f_1(t)} {f_2(u)}}{f_1(u) (f_2(t)-f_2(s)) - f_1(t) (f_2(s) + f_2(u)) + f_1(s) (f_2(t) + f_2(u))},\\ \nonumber
\delta(t,s,u)&=&  \frac{{f_1(s)} {f_2(t)}-{f_1(t)}{f_1(u)}}{f_1(u)(f_2(t)-f_2(s)) - f_1(t) (f_2(s) + f_2(u)) + f_1(s) (f_2(t) + f_2(u))},\\ 
\label{e.weights}
\end{eqnarray}
where
\[f^T(x)=\left(-\frac{x}{\alpha+\beta \gamma x^{-1+\gamma}},-\frac{x^\gamma}{\alpha+\beta \gamma x^{-1+\gamma}},-\frac{\beta \gamma \
x^\gamma \log x}{\alpha+\beta \gamma x^{\gamma-1}}\right).\]
Plugging (\ref{e.weights}) into the third component the function
\begin{eqnarray*}
x_3(t,s,u)= (1-\lambda(t,s,u)-\delta(t,s,u))\, f_3(t)+\lambda(t,s,u) f_3(s)-\delta(t,s,u) f_3(u)
\end{eqnarray*}
has to be maximized subject to $t,s,u \in \mathcal{X}_{netOD}=[0,0.45]$. The maximum is reached at $t^*=0.06, s^*=0.27, u^*=0.45$. The weights are them obtained from Equation (\ref{e.weights}),
\begin{eqnarray*}
1-\lambda(t^*,s^*,u^*)-\delta(t^*,s^*,u^*)=0.41, \quad\lambda(t^*,s^*,u^*)=0.40, \quad \delta(t^*,s^*,u^*)=0.19.
\end{eqnarray*}

Thus, the c-optimal designs  before, $\xi^{netOD}_{\gamma}$, and after, $\xi_{\gamma}^{Dose}$, the transformation, are given in Table \ref{t.new}. This is the best design to estimate the parameter $\gamma$. Proceeding in a similar way the best designs for estimating $\alpha$ and $\beta$ are computed. Since the c-optimal design for estimating $\beta$ is a singular two--point design, and therefore the D--efficiency is zero, the c-efficiencies of the D-optimal design are shown in Table \ref{t.new} instead of the D--efficiencies. The c-optimal designs for the parameters $\alpha$ and $\gamma$ have the same support points but different weights, this implies that the D-optimal design has a efficiency greater than 80\% for estimating $\gamma$ but for estimating $\alpha$ that efficiency decreases and it is close to 50\%. The c-efficiency of the D-optimal design for estimating parameter $\beta$ is too low due to fact that the c-optimal design is singular.

\subsection{Space--filling  designs with more than 3 points}\label{s.spfil}

Generally, the experimenters do not like designs with few and extreme points. For instance, the design used by \cite{Rein12} was the collection of equidistant points between 0.2 and 8 using steps of 0.5, that is
$$\xi_E=\{ 0.2, 0.7, 1.2, 1.7, 2.2, 2.7, 3.2, 3.7, 4.2, 4.7, 5.2, 5.7, 6.2, 6.7, 7.2, 7.7\}$$
for the doses. Using equation (\ref{e.radmodel}) the design is transformed into
$$\{0.024, 0.077, 0.12, 0.16, 0.19, 0.22, 0.24, 0.26, 0.28, 0.30, 0.32, 0.34, 0.35, 0.37, 0.38, 0.39\}$$ 
for $netOD$. The D-efficiency of this design is about 51\%. But this design can be very much improved still keeping the requirements of the practitioners. If an exact design with a number of  points, say $n$, is searched then three different, but replicated in some way, points are going to be found always. If more than three different points are wanted then the search has to be forced to a sequence of points following some particular rule, such as an arithmetic or geometric sequence. It is worthy to explain here that this has nothing to do with sequential or adaptive designs. That is the reason they are called space--filling designs since they try to be spread along the design space. In these cases, the D-optimal design can be used as a reference measure of the goodness of the space--filling design considered. \cite{Lop02} optimized different types of sequences according to D-optimality, including arithmetic, geometric, harmonic and an arithmetic inverse of the trend model. In this section, D-optimal space--filling designs are computed and compared in order to analyze these designs because we knew by personal communication with our physicists collaborators that there was particular interest in them.

The target are exact designs of the type
$$ \xi_{n}=\{Dose_{1},Dose_{2}, ... ,Dose_{n}\},$$
after a transformation from a design on $netOD$.

Each of the two sequences may follow some pattern, e.g. arithmetic or geometric rules. The FIM is computed as:
$$ M(\xi_{n},\theta)=\frac{1}{n} \sum_{i}I(Dose_{i},\theta)= \frac{1}{n} \sum_{i}f(netOD_{i})f^T(netOD_{i}). $$

Space--filling designs  of size $n = 6$ will be considered in this paper, although the main idea remains for any sample size. Arithmetic and geometric sequences are being considered in the sense explained in what follows. In all the cases appropriate interesting efficiencies will be obtained.

\subsubsection*{Arithmetic sequences}

Taking into account the last point of the design space $\mathcal{X}_{netOD}=[0, b]$ is always in the support of the D-optimal design, the arithmetic sequence will be forced to end at the right extreme of the interval,
$$ b (1 - r), b \left(1 - r\frac{n-2}{n-1}\right), ...,  b \left(1 - r\frac{2}{n-1}\right),  b \left(1 - r\frac{1}{n-1}\right), b; \; r\in (0,1).$$

Just $r$ will be free and it will be optimized. The FIM for this sequence assuming equal weights at each point, i.e. $1/n$, is
$$ M(\xi_{n},\theta)=\frac{1}{n} \sum_{i=1}^{n}I(Dose_{i},\theta)= \frac{1}{n} \sum_{i=1}^{n}f\left(b \left(1 - r\frac{n-i}{n-1}\right)\right)f^T\left(b \left(1 - r\frac{n-i}{n-1}\right)\right).$$

\begin{figure}[h]
	\begin{center}
		\includegraphics[scale=0.4]{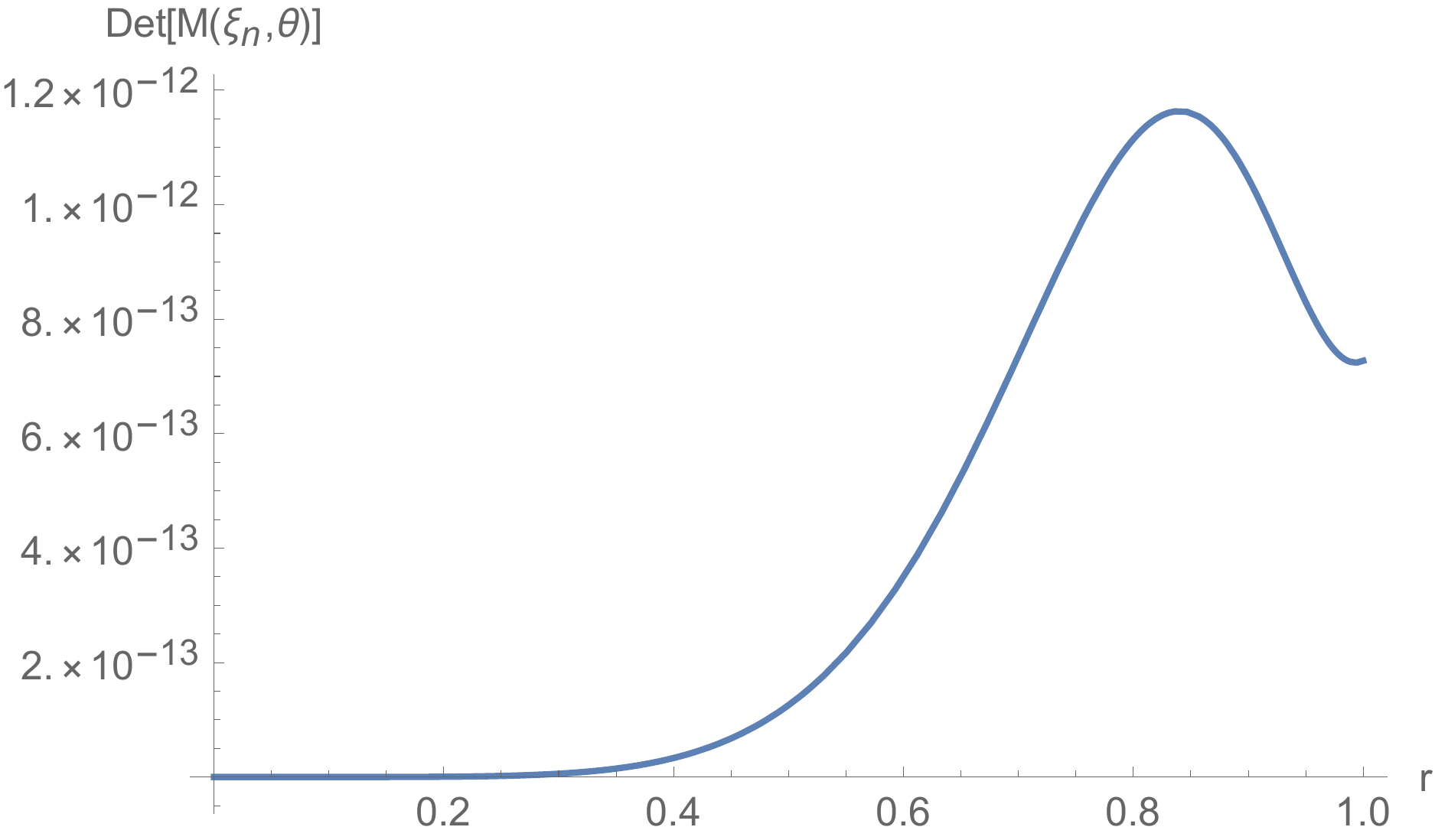}\includegraphics[scale=0.4]{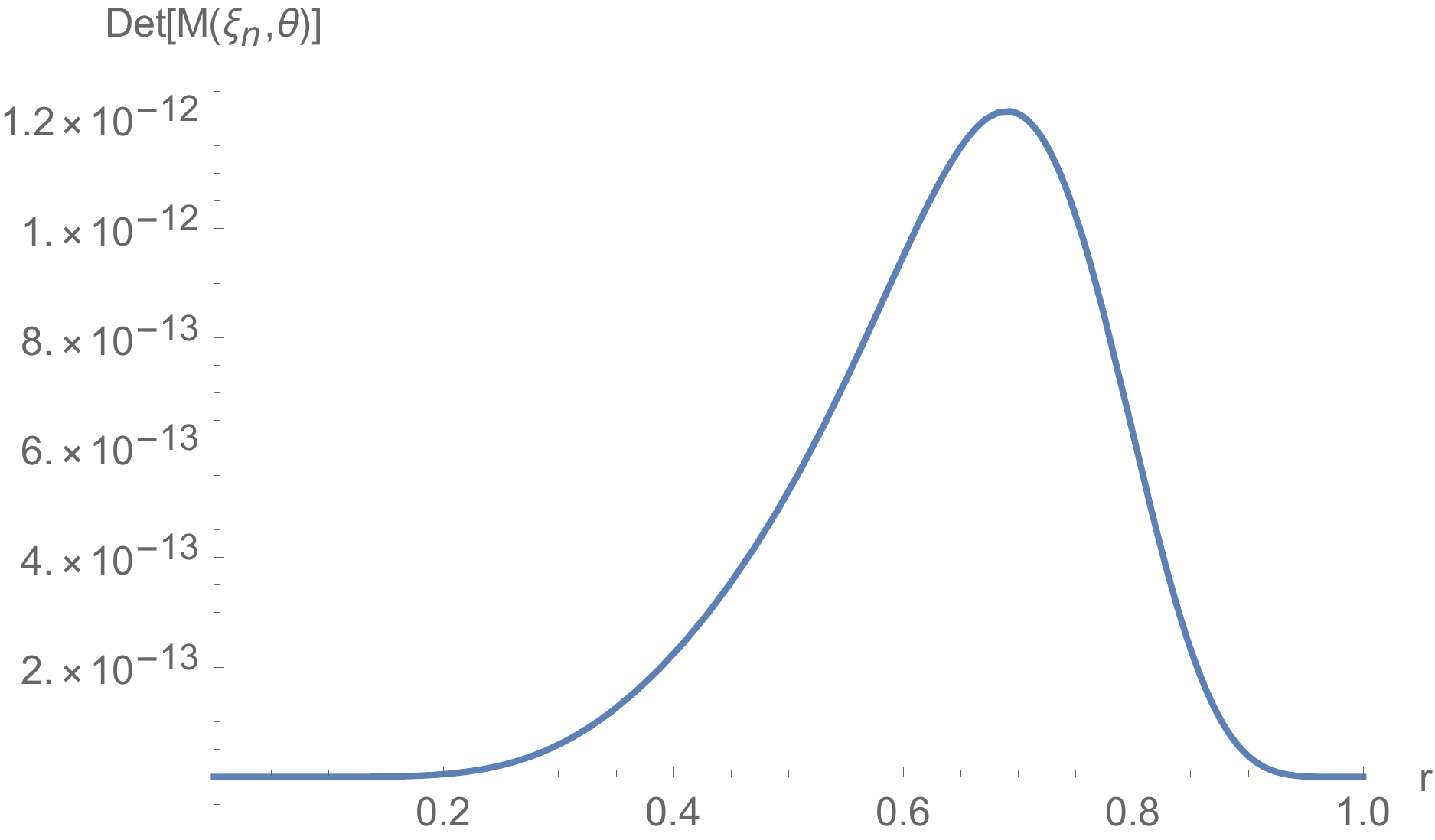}
	\end{center}
	\caption{Determinant of the FIM for an arithmetic (left) and a geometric sequence (right) on $netOD$ in function of $r$.}\label{f.sequen}
\end{figure}

A ratio of $r^*=0.84$ maximizes the determinant as it is shown in Figure \ref{f.sequen} (left). The arithmetic D-optimal sequences on $netOD$ and $Dose$, $\xi^{Ar}_{D}$, as well as its D-efficiency are shown in Table \ref{t.new}.

\subsubsection*{Geometric sequences}

We consider geometric sequences starting at the last point, $b=0.45$,  going backwards,
$$ r^{n-1}b, ..., r^{2}b, rb, b.$$
The ratio, $r$, has to be optimized. The FIM for a design of this type is
$$ M(\xi_{n},\theta)=\frac{1}{n} \sum_{i}I(Dose_{i},\theta)= \frac{1}{n} \sum_{i=1}^{n}f(r^{n-i}b)f^T(r^{n-i}b).$$

Figure \ref{f.sequen} (right) shows the ratio $r^*=0.69$ maximizes the determinant. The geometric D-optimal designs (on $netOD$ and $Dose$) and its D-efficiency are shown in Table \ref{t.new}. In spite of the designs for both sequences are different the efficiencies are quite similar and high. The reason could be related to the presence of the end-point of the design space, 0.45, in these designs and that the first support point of the sequences, 0.07 is close to the first support point of the D-optimal design, 0.09.



\section{Optimal designs for estimating the right dose}\label{sec:cal}

In previous sections the main concern of the experimental design was the precise estimation of the parameters of the model. But the main interest in calibration, from the point of view of the practitioners, is the precise prediction (calibration) of the explanatory variable. Thus, optimal designs in the explanatory variable should be computed to minimize the variance of the prediction in this variable. As mentioned in the introduction, \cite{Fra04} computed optimal designs for inverse prediction in calibration models, in the particular case where the function of the model is known and invertible. They presented two criteria, G$_I-$optimality and V$_I-$optimality in this paper. The aims of these criteria are the same that G- and V-optimality, but for inverse prediction. In this section, we adapt these criteria to the problem faced in this work.

Taking into account that the explanatory variable $Dose$ is modelled by the function $\mu(netOD,\theta)$, the variance of the prediction of $Dose$ given a value of $netOD$ is
$$Var(\hat{Dose})=\frac{\partial\mu(netOD,\theta)}{\partial\theta^T}M^{-1}(\xi_D,\theta)\frac{\partial\mu(netOD,\theta)}{\partial\theta}.$$

Then, from this definition of the variance of the explanatory variable we can define the following criteria,
\begin{eqnarray*}
\Phi_{G_I}(\xi)&=&\max_{netOD \in \mathcal{X}_{netOD}} Var(\hat{Dose})\\
\Phi_{V_I}(\xi)&=&\frac{1}{b}\int_a^b Var(\hat{Dose})\,d(netOD),
\end{eqnarray*}
where $b$ is the range of $\mathcal{X}_{netOD}$. Both criteria can be optimized in $netOD$ and the optimal design in this variable will be transformed to the optimal design in $Dose$ following the procedure used in previous sections.

\subsection{Computation of G$_I-$optimal designs}

For computing the G$_I-$optimal design the following algorithm is proposed.
\begin{description}
\item[\textbf{\emph Algorithm}]
\item[Step 1. ]Select an initial design $\xi_0$ supported on at least three different points.

\item[Step 2. ]Let $\xi_s$ be the design obtained at step $s$. Determine $$netOD_s=\arg \max_{netOD \in \mathcal{X}_{netOD}} \frac{\partial\mu(netOD,\theta)}{\partial\theta^T}M^{-1}(\xi_s,\theta)\frac{\partial\mu(netOD,\theta)}{\partial\theta}.$$

\item[Step 3. ]Let  $\xi_{s+1}=(1-\alpha_s)\xi_s+\alpha_s\xi_{netOD_s},$ where $\xi_{netOD_s}$ is a one-point design with $\alpha_s=1/(s+1)$ or some optimized step satisfying the conditions $\alpha_s \rightarrow 0$, $\sum_s \alpha_s =\infty$.

\item[Step 4. ]If $$1+\frac{\max_{netOD \in \mathcal{X}_{netOD}} \frac{\partial\mu(netOD,\theta)}{\partial\theta^T}M^{-1}(\xi_{s+1},\theta)\frac{\partial\mu(netOD,\theta)}{\partial\theta}}{\Phi_{G_I}(\xi_{s+1})}>\delta,$$ where $\delta$ is the given efficiency bound, then STOP. Otherwise, set
$s+1\leftarrow s+2$ and go to Step 2.
\end{description}

Using this algorithm the G$_I-$optimal design obtained on $netOD$ and the transformed design are given in Table \ref{t.new}, as well as its D-efficiency.

\subsection{Computation of V$_I-$optimal designs}

For computing the V$_I-$optimal designs, firstly the sensitivity function is calculated in order to define the algorithm. The sensitivity function for V$_I-$criterion is

$$
\psi(z,\xi)=\int_{0}^b\left(\frac{\partial\mu}{\partial\theta^T}M^{-1}(\xi_D,\theta)\frac{\partial\mu}{\partial\theta}-\right.
$$

$$
\left. -\left(\frac{\partial\mu}{\partial\theta}\right)M^{-1}(\xi_D,\theta)\frac{\partial\eta(z,\theta)}{\partial\theta}\frac{\partial\eta(z,\theta)}{\partial\theta^T}M^{-1}(\xi_D,\theta)\frac{\partial\mu}{\partial\theta^T}\right)\, d(netOD),
$$
where $\mu\equiv \mu(netOD,\theta)$  for  space saving.
\begin{description}
\item[\textbf{\emph Algorithm}]
\item[Step 1. ]Select an initial design $\xi_0$ supported on at least three different points.

\item[Step 2. ]Let $\xi_s$ be the design obtained at step $s$ and compute $$netOD_s=\arg \max_{netOD \in \mathcal{X}_{netOD}} \psi(z,\xi_s).$$

\item[Step 3. ] Define $\xi_{s+1}=(1-\alpha_s)\xi_s+\alpha_s\xi_{netOD_s},$ where $\xi_{netOD_s}$ is a one-point design with $\alpha_s=1/(s+1)$ or some optimized step satisfying the conditions $\alpha_s \rightarrow 0$, $\sum_s \alpha_s =\infty$.

\item[Step 4. ]If $$1+\frac{\max_{netOD \in \mathcal{X}_{netOD}} \psi(netOD,\xi_{s+1})}{\Phi_{V_I}(\xi_{s+1})}>\delta,$$ where $\delta$ is the
given efficiency aimed, then STOP. Otherwise, set
$s+1\leftarrow s+2$ and go to Step 2.
\end{description}

With this algorithm the V$_I-$optimal design obtained on $netOD$, the transformed design and its D-efficiency are shown in Table \ref{t.new}. It is remarkable the difference between the D-efficiency of the V$_I-$ and G$_I-$optimal design. Thus, while the efficiency for the estimation of the parameters of the model is about 60\% with G$_I-$optimality, the V$_I-$optimal design is a suitable design both for estimation and for prediction.       

\subsection{A note on the convergence of the algorithms for G$_I$ and V$_I$--optimality}

As seen in Section \ref{sec.inv} the calibration model takes to a transformed model that is heteroscedastic but the variance of the predictions of the explanatory variable are considered in these two criteria. In order to make it more clear a general situation is considered. Let $f(x)=g(x)/w(x)$ as in (\ref{e.transfim}), where the three functions depend also on $\theta$. The parameters are omitted here for brevity taking also into account the nominal values will be provided from the beginning. Then the information matrix of a design is built with $f$, 
\begin{eqnarray*}
M(\xi,\theta)=\sum_{x\in\mathcal{X}} f(x)f^T(x)\xi(x). 
\end{eqnarray*}

V$_I$--optimality is again a type of V--optimality with a different measure, in particular the Lebesgue measure multiplied by $w^2(x)$,
 \begin{eqnarray*}
\Phi_{V_I} (M(\xi,\theta))=\int_\mathcal{X} g^T(x)M_g^{-1}(\xi) g(x) w^2(x) dx,
\end{eqnarray*}
and therefore the convergence of the algorithm is known.

For G$_I$--optimality the convergence is not so simple. The G$_I$--optimality criterion is defined as
\begin{eqnarray*}
\Phi (M(\xi,\theta))&=&\max_{x\in\mathcal{X}} g^T(x)M^{-1}(\xi) g(x)\\
&=&\max_{x\in\mathcal{X}} w^2(x)f^T(x)M^{-1}(\xi) f(x)\\
&=&\max_{x\in\mathcal{X}} w^2(x)d(x,\xi),
\end{eqnarray*}
where $d(x,\xi)=f^T(x)M^{-1}(\xi) f(x)$ is proportional to the variance of the prediction of the response as usual. This criterion is different from G--optimality in the factor $w^2(x)$. It is convex and if $w^2(\cdot)$ is continuous in the compact set $\mathcal{X}$  then there exists an optimal design $\xi^\star$ with $\Phi^\star = \Phi (\xi^\star)>-\infty$. Our empirical results seem to lead to an optimal design.

\subsection{Space--filling designs}
Following the procedure of Section \ref{s.spfil} designs with 6 points have been computed also for these criteria. They are shown in Table \ref{t.new} with their efficiencies with respect to the criterion considered, G$_I-$optimality and V$_I-$optimality. We can see how these space--filling designs are better for V$_I-$optimality than for G$_I-$optimality, where the efficiencies are lower than 40\%.

\begin{table}[h]
\centering
\begin{tabular}{lcc|cc|r}\hline
& \multicolumn{2}{c}{$netOD$}  &\multicolumn{2}{c}{$Dose$}  &Efficiency \\
Design& \multicolumn{2}{c}{Support points (weights)}  &\multicolumn{2}{c}{Support points (weights)}  & \% \\ \hline
$\xi_{D}$  & 0.09 (1/3) & 0.27 (1/3)&  0.80 (1/3)& 3.90 (1/3)&  100\\
$\tilde{\xi}_{D}$ & 0.13 (1/3)& 0.33 (1/3)&  1.33 (1/3)& 5.54 (1/3)&  87\\
$\xi_{\alpha}$ & 0.06 (0.78)& 0.27 (0.16)&  0.54 (0.78)& 4.00 (0.16)&  53\\
$\xi_{\beta}$ & 0.29 (0.48)& &  4.45 (0.48)& &  0.3\\
$\xi_{\gamma}$  & 0.06 (0.41)& 0.27 (0.40)&  0.54 (0.41)& 4.00 (0.40)&  81\\
$\xi_{G_I}$  & 0.13 (0.06)& 0.33 (0.30)&  1.33 (0.06)& 5.54 (0.30)&  59\\
$\xi_{V_I}$  & 0.09 (0.19)& 0.29 (0.46)&  0.84 (0.19)& 4.41 (0.46)&  93\\
$\xi_{D}^{Ar}$ & 0.07 \hspace{0.1cm} 0.15& 0.22 \hspace{0.1cm} 0.30  \hspace{0.1cm} 0.37&0.60 \hspace{0.1cm} 1.50 &2.80 \hspace{0.1cm} 4.60 \hspace{0.1cm} 6.90 & 84\\
$\xi_{D}^{Ge}$ & 0.07 \hspace{0.1cm} 0.10 &0.15 \hspace{0.1cm} 0.21 \hspace{0.1cm} 0.31 &0.60 \hspace{0.1cm} 0.90 &1.50 \hspace{0.1cm} 2.60 \hspace{0.1cm} 4.90 &85 \\
$\xi_{G_I}^{Ar}$ & 0.23 \hspace{0.1cm} 0.27 & 0.31 \hspace{0.1cm} 0.36 \hspace{0.1cm} 0.41 &  2.90 \hspace{0.1cm} 3.90 & 5.00 \hspace{0.1cm} 6.40 \hspace{0.1cm} 8.10 &  37  \\
$\xi_{G_I}^{Ge}$ & 0.22 \hspace{0.1cm} 0.26 & 0.30 \hspace{0.1cm} 0.34 \hspace{0.1cm} 0.39 &  2.80 \hspace{0.1cm} 3.60 & 4.50 \hspace{0.1cm} 5.80 \hspace{0.1cm} 7.60 & 29 \\
$\xi_{V_I}^{Ar}$ & 0.10 \hspace{0.1cm} 0.17 & 0.24 \hspace{0.1cm} 0.31 \hspace{0.1cm} 0.38   & 0.90 \hspace{0.1cm} 1.80 & 3.20 \hspace{0.1cm} 4.90 \hspace{0.1cm} 7.10 &  83 \\
$\xi_{V_I}^{Ge}$ & 0.13 \hspace{0.1cm} 0.17 & 0.21 \hspace{0.1cm} 0.27 \hspace{0.1cm} 0.35 &  1.30 \hspace{0.1cm} 1.80 & 2.60 \hspace{0.1cm} 3.90 \hspace{0.1cm} 6.20 &  73 \\ \hline
\end{tabular}
\caption{Optimal designs (support points and weights within parenthesis when it is the case). The last point of all the designs, either 0.45 or 10, is ommited. The efficiencies of the approximate designs are computed with respect to the D-optimal design except the efficiencies for the c-optimal designs that are the c-efficiencies of the D-optimal design. For the sequences the efficiencies are computed with respect to the corresponding criterion, either D, V$_I$ or G$_I$, for the optimal approximate design}\label{t.new}
\end{table}

\section{Concluding remarks}
By using the Inverse Function Theorem, optimal designs were computed on the dependent variable for estimating the parameters of the model and for the prediction of the independent variable considering a dosimetry model. From the perspective of the estimation of the parameters, the D-optimal design was computed directly on the response variable and then it was transformed into a design on the explanatory variable. This is not the proper design to be computed and may displayed an important loss of efficiency as is the case in our example with respect to the right one. The transformed model actually becomes heteroscedastic. Optimal designs for estimating each parameter of the model were also computed. This allowed to measure how efficient was the D-optimal design for estimating each of them, displaying a good efficiency for estimating parameter $\gamma$ but neither for $\alpha$ nor for $\beta$.

Taking into account that this model has calibration purposes, the G$_I-$ and V$_I-$optimal designs have been computed in order to optimize the inverse prediction. Moreover, algorithms for computing them are provided in this paper.

Since three--point designs may be not acceptable from a practical point of view six different points were imposed to be in the design forcing them to follow a regular sequence. In particular, arithmetic and geometric sequences were considered. All of them were more efficient than the sequence used by the researchers. In particular, Table \ref{t.exp} shows the efficiencies of $\xi_E$ with respect to the computed designs in this paper. This stresses the importance of using a good design.

\begin{table}[h]
\centering
\begin{tabular}{lcccccc}\hline
Criterion	&D&	G$_I$	&V$_I$	&c$_1$	&c$_2$	&c$_3$\\
Efficiency	&0.51	&0.07	&0.34	&0.19	&0.18	&0.24\\  \hline
\end{tabular}
\caption{Efficiencies of the experimental design used in practice with respect to different criteria}\label{t.exp}
\end{table}

Arithmetic and geometric sequences can be considered on the $Dose$ as well. In particular, the arithmetic sequence woud be
$$B \left(1 - r\frac{n-i}{n-1}\right), \; i=1,...,n,\; r\in (0,1).$$
In order to compute the FIM and the optimal design these points have to be transformed into the corresponding sequence $netOD_i, \; i=1,...,n$. For that the following equation needs to be solved numerically for the corresponding nominal values ,
$$B \left(1-\frac{n-i}{n-1} r \right)=\alpha netOD_i+\beta netOD_i^\gamma.$$

Assuming again the last point of the design space is always in the support of the optimal design the geometric sequences are now considered starting at the last point, $B$, and going backwards,
$$ r^{n-1}B, ..., r^{2}B, rB, B.$$
This sequence has to be transformed in a sequence on variable $netOD$ solving numerically for the assumed nominal values the equation,
$$B r^{n - i} = \alpha netOD + \beta netOD^{\gamma}$$

The main contribution of this work is to establish the methodology for computing optimal designs when the function of the model is given as a function of the response variable and there is not a closed-form available for its inverse. This situation is common with calibration models as the dosimetry model considered in this work. Thus, the optimal designs computed cover different aims from the estimation of the parameters of the model to the prediction of the proper dose to be applied to a patient. There is a huge improvement of the latter if a specific dose is determined for each particular patient according to his or her different personal or clinical features. This may be done using the ideas of \cite{LoGa04} or \cite{MaToLo07}.

\section*{Acknowledgements}
The authors want to thank Dr. Ramos-Garc\'ia for his collaboration introducing us with the model considered in the paper. This work was sponsored by Ministerio de Economía y Competitividad MTM2016-80539-C2-1-R and by Consejería de Educación, Cultura y Deportes of Junta de Comunidades de Castilla-La Mancha and Fondo Europeo de Desarrollo Regional SBPLY/17/180501/000380.

\bibliographystyle{plainnat}
\bibliography{mybibfile}

\begin{thebibliography}{20}
\providecommand{\natexlab}[1]{#1}
\providecommand{\url}[1]{\texttt{#1}}
\expandafter\ifx\csname urlstyle\endcsname\relax
  \providecommand{\doi}[1]{doi: #1}\else
  \providecommand{\doi}{doi: \begingroup \urlstyle{rm}\Url}\fi

\bibitem[Amo-Salas et~al.(2016)Amo-Salas, Jim\'enez-Alc\'azar, and
  L\'opez-Fidalgo]{Amo16}
M.~Amo-Salas, A.~Jim\'enez-Alc\'azar, and J.~L\'opez-Fidalgo.
\newblock Optimal designs for implicit models.
\newblock In A.C.~Atkinson J.~Kunert, Ch.H.~Muller, editor, \emph{mODa 11 –-
  Advances in Model-Oriented Design and Analysis}, pages 11--18, Heidelberg,
  2016. Springer.

\bibitem[Atkinson et~al.(2007)Atkinson, Donev, and Tobias]{Atk07}
A.C. Atkinson, A.N. Donev, and R.D. Tobias.
\newblock \emph{Optimum Experimental Designs, with SAS}.
\newblock Oxford Statistical Science Series, 2007.

\bibitem[Bartroff(2011)]{Ba11}
J.~Bartroff.
\newblock A new characterization of elfving's method for high dimensional
  computation.
\newblock \emph{Journal of Statistical Planning and Inference}, 142\penalty0
  (4):\penalty0 863--871, 2011.

\bibitem[Biedermann et~al.(2011)Biedermann, Bissantz, Dette, and
  Jones]{Biederman11}
S.~Biedermann, N.~Bissantz, H.~Dette, and E.~Jones.
\newblock Optimal designs for indirect regression.
\newblock \emph{Inverse Problems}, 27:\penalty0 1--19, 2011.

\bibitem[Elfving(1952)]{Elf52}
G.~Elfving.
\newblock Optimum allocation in linear regression theory.
\newblock \emph{Ann. Math. Stat.}, 23\penalty0 (2):\penalty0 255--262, 1952.
\newblock \doi{10.1214/aoms/1177729442}.

\bibitem[Fedorov and Hackl(1997)]{Fed97}
V.~V. Fedorov and P.~Hackl.
\newblock \emph{Model-Oriented Design of Experiments. Lecture Notes in
  Statistics.}
\newblock Springer, 1997.

\bibitem[Francois et~al.(2004)Francois, Govaerts, and Boulanger]{Fra04}
N.~Francois, B.~Govaerts, and B.~Boulanger.
\newblock Optimal designs for inverse prediction in univariate nonlinear
  calibration models.
\newblock \emph{Chemometrics and Intelligent Laboratory Systems}, 74:\penalty0
  283--292, 2004.
\newblock \doi{10.1016/j.chemolab.2004.05.005}.

\bibitem[Harman and Jurik(2008)]{ha08b}
Radoslav Harman and Tomas Jurik.
\newblock Computing c-optimal experimental designs using the simplex method of
  linear programming.
\newblock \emph{Computational Statistics \& Data Analysis}, 53\penalty0
  (2):\penalty0 247--254, 2008.
\newblock \doi{10.1016/j.csda.2008.06.023}.

\bibitem[Higueras et~al.(2019)Higueras, Howes, and L\'opez-Fidalgo]{mH19}
M.~Higueras, A.~Howes, and J.~L\'opez-Fidalgo.
\newblock Optimal experimental design for cytogenetic dose-response calibration
  curves.
\newblock \emph{International Journal of Radiation Biology (submitted)}, 2019.

\bibitem[Kiefer and Wolfowitz(1960)]{Kie60}
J.~Kiefer and J.~Wolfowitz.
\newblock The equivalence of two extremum problems.
\newblock \emph{Canad. J. Math.}, 12:\penalty0 363--366, 1960.
\newblock \doi{10.4153/CJM-1960-030-4}.

\bibitem[Kitsos(1992)]{Kit92}
C.~P. Kitsos.
\newblock \emph{Quasi-Sequential Procedures for the Calibration Problem}, pages
  227--231.
\newblock Physica-Verlag HD, 1992.
\newblock \doi{10.1007/978-3-642-48678-4_27}.

\bibitem[L\'opez-Fidalgo and Garcet-Rodr\'{i}guez(2004)]{LoGa04}
J.~L\'opez-Fidalgo and S.~Garcet-Rodr\'{i}guez.
\newblock Optimal experimental designs when some independent variables are not
  subject to control.
\newblock \emph{J. Am. Statist. Assoc.}, 99\penalty0 (468):\penalty0
  1190--1199, 2004.

\bibitem[L\'opez-Fidalgo and Rodr\'{\i}guez-D\'{\i}az(2004)]{Lop04}
J.~L\'opez-Fidalgo and J.M. Rodr\'{\i}guez-D\'{\i}az.
\newblock Elfving's methods for m-dimensional models.
\newblock \emph{Metrika}, 59\penalty0 (3):\penalty0 235--244, 2004.
\newblock \doi{10.1007/s001840300281}.

\bibitem[L\'opez-Fidalgo and Wong(2002)]{Lop02}
J.~L\'opez-Fidalgo and W.~K. Wong.
\newblock Design issues for the michaelis-menten model.
\newblock \emph{J. Theoret. Biol.}, 215:\penalty0 1--11, 2002.
\newblock \doi{10.1006/jtbi.2001.2497}.

\bibitem[Mart{\'\i}n-Mart{\'\i}n et~al.(2007)Mart{\'\i}n-Mart{\'\i}n, Torsney,
  and L\'opez-Fidalgo]{MaToLo07}
R.~Mart{\'\i}n-Mart{\'\i}n, B.~Torsney, and J.~L\'opez-Fidalgo.
\newblock Construction of marginally and conditionally restricted designs using
  multiplicative algorithms.
\newblock \emph{Computational Statistics and Data Analysis}, 51:\penalty0
  5547--5561, 2007.

\bibitem[Osborne(1991)]{Osb91}
C.~Osborne.
\newblock Statistical calibration: a review.
\newblock \emph{International Statistical Review}, 59\penalty0 (3):\penalty0
  309--336, 1991.
\newblock \doi{10.2307/1403690}.

\bibitem[Pazman(1986)]{Paz86}
A.~Pazman.
\newblock \emph{Foundations of optimum experimental design}.
\newblock D. Reidel publishing company, 1986.

\bibitem[Ramos-Garc\'{\i}a and P\'erez-Azor\'{\i}n(2013)]{Ram13}
L.~I. Ramos-Garc\'{\i}a and J.F. P\'erez-Azor\'{\i}n.
\newblock Improving the calibration of radiochromic films.
\newblock \emph{Medical Physics}, 40\penalty0 (7):\penalty0 17--26, 2013.
\newblock \doi{10.1118/1.4811238}.

\bibitem[Reinhardt et~al.(2012)Reinhardt, Hillbrand, Wilkens, and
  Assmann]{Rein12}
S.~Reinhardt, M.~Hillbrand, J.J. Wilkens, and W.~Assmann.
\newblock Comparison of gafchromic ebt2 and ebt3 films for clinical photon and
  proton beams.
\newblock \emph{Medical Physics}, 39\penalty0 (8):\penalty0 5257--5262, 2012.

\bibitem[Whittle(1973)]{Whi73}
P.~Whittle.
\newblock Some general points in the theory of optimal experimental design.
\newblock \emph{Journal Royal Statistical Society, Ser. B}, 1:\penalty0
  123--130, 1973.

\end{thebibliography}

\end{document}